\documentclass[aps,epsfig,rotate,showpacs]{revtex4}
\usepackage{epsfig}
\usepackage{graphicx}
\usepackage{amsmath}

\newcommand{\beq}{\begin{eqnarray}}
\newcommand{\eeq}{\end{eqnarray}}
\begin{document}                

\title{A General Method for Complete Population Transfer in Degenerate Systems}
\author{Jiangbin Gong and Stuart A. Rice}
\affiliation{Department of Chemistry and The James Franck Institute,\\
The University of Chicago, Chicago, Illinois 60637}
\date{\today}

\begin{abstract}
A simple theoretical solution to the design of a control field
that generates complete population transfer from an initial state, via $N$ nondegenerate intermediate states,
to one arbitrary member of $M$ ($M\leq N$) degenerate states
is constructed. The full control field exploits an $(M+N-1)$-node null adiabatic state, created by designing the relative phases and
amplitudes of the component fields that together make up the full field.
The solution found is universal in the sense that it does not depend on the exact 
number of the unwanted degenerate states or their properties.
The results obtained suggest that a class of multi-level quantum systems with degenerate states can be completely controllable,
even under extremely strong constraints, {\it e.g.}, never populating a Hilbert subspace
that is only a few dimensions smaller than the whole Hilbert space.
\end{abstract}
\maketitle

\section{Introduction}

Coherent control, {\it i.e.}, controlling atomic and molecular processes via quantum interference effects 
induced by external fields,
has attracted great theoretical and experimental interest \cite{ricebook,rice01,brumerbook}.  
In particular, our ability to manipulate atomic and molecular excitation and therefore drive the microscopic motion to generate an 
arbitrarily selected target state
is one key prerequisite for
quantum information processing \cite{qcbook} in atoms and molecules.

One formal mathematical problem associated with coherent control is determining
whether or not  complete controllability is achievable.
If, for a given system, it can be proved that there exists 
a control field (possibly composed of a number of component fields)
that can yield complete population transfer to an arbitrary target state 
within a finite time, then the system is called completely
controllable. For such systems, the primary task of coherent control studies is to design methods that are
conceptually as simple as possible and also experimentally feasible.
If the formal complete controllability of a quantum system is
not established in general,
then it is of great theoretical interest to 
construct specific control fields that can achieve, say,
complete population transfer between two particular quantum states.

Some existence theorems
that establish conditions for complete controllability of a quantum system
have been proved \cite{huang1,rama,tersigni}.  The Huang-Tarn-Clark theorem \cite{huang1} is believed to be
the strongest result,
but it applies only to systems with discrete and nondegenerate states.  Ramakrishna {\it et al.} showed that for a quantum system with a Hilbert space of dimension $L$,
the necessary and sufficient condition for complete controllability is that the field-free Hamiltonian and the interaction Hamiltonian induced by a
control field
generate a Lie algebra of dimension $L^{2}$ \cite{rama}. This criterion is applicable to multi-level systems that involve degenerate states \cite{shah,rabitz}, but the required
computations to generate the commutators of the Lie algebra structure can be demanding for large $L$,  vary drastically from system to system,
and provide no hint as to how a control field can be designed.  
As such, the possibility
for complete controllability
in a variety of degenerate systems is still unknown and
it is widely accepted that
coherent control in degenerate systems is more difficult than in
nondegenerate systems \cite{kobrakpra,shapiro,legare,gongjcp2}.  

Even more challenging is to establish the complete controllability of a quantum system subject to restrictions on the population transfer pathway
along which the system may evolve.
Shapiro and Brumer \cite{shapirojcp} have shown that there is often a loss of controllability if the whole Hilbert space of a quantum system
is partitioned and one wishes to generate an evolution that does not pass through some particular Hilbert subspace.
Sol\'{a} {\it et al.} \cite{solapra}  have proposed an optimal control theory to maximize the final population in the target state and
minimize a time-integral of the population
in all the unwanted states.  Neither of these two previous studies has provided conditions for
a class of quantum systems subject to strong constraints
to be completely controllable.  

Encouraged by recent progress in extending
the stimulated Raman adiabatic passage (STIRAP) \cite{bergmann} method and 
its variations \cite{gongjcp2,gongjcp1,gongjcp3,gongjcp4,unanyan0,bergmannprl,unanyan,kis,karpati}
for coherent control, we address the controllability of population transfer
to one of $M$ degenerate states in a system with $N\geq M$ nondegenerate intermediate states by invoking
a general adiabatic passage scheme.
In particular,
we demonstrate that, under certain conditions,
it is possible to realize complete population transfer from an initial state, via $N$ non-degenerate intermediate states, to 
one arbitrary member of a set of $M$ ($M\leq N$) degenerate states, without ever populating any states other than the initial state and the target state.
The solution is surprisingly simple and general in the sense that realizing such a population transfer pathway may not require our knowledge of
the exact number of degeneracies or the properties of the $M$ degenerate states except for the target state.
The solution can also be easily adapted for the creation of 
an arbitrary superposition of the $M$ degenerate states.
The results we have obtained
strongly  suggest that a class of multi-level quantum systems with degenerate states should be completely controllable.
The results also imply that complete controllability in a class of quantum systems
may still be possible even when extremely strong constraints are applied,
 {\it e.g.}, never populating a Hilbert subspace
that is only dimension-two smaller than the whole Hilbert space.    

The central idea of our approach is to manipulate the component fields of the full control field
to create  
an $(M+N-1)$-node null eigenstate of the system dressed by the 
fields. This multi-node null eigenstate is designed to fully 
correlate with the initial state at early times and
then evolve to the target state
as the component fields that have specified 
relative phases and amplitudes are turned on and off in a particular order.
This scheme is a significant extension of our previous
adiabatic passage method 
for the realization of complete control of 
the population transfer branching ratio between two degenerate states \cite{gongjcp4}.
In addition, both the present study and our previous work \cite{gongjcp4}
demonstrate a powerful marriage between  weak-field coherent phase control and 
traditional adiabatic passage techniques \cite{bergmann} that are insensitive to the relative phases of the control fields.

This paper is organized as follows. In Sec. II we describe our model system.
We then present in Sec. III our adiabatic passage scheme for the complete
population transfer from an initial state to one arbitrary member of $M$ degenerate states.
Simple computational examples that support our theoretical results  are 
described in Sec. IV.
We discuss our results and conclude this paper in Sec. V.

\section{The model system}
Consider a multi-level quantum system that has a nondegenerate
 initial state (denoted $|0\rangle$), and $M$ degenerate orthogonal states (denoted $|f_{1}\rangle$,  $|f_{2}\rangle$,
 $\cdots$, $|f_{M}\rangle$).  Without loss of generality we denote one arbitrary member of the $M$ degenerate states 
that we wish to transfer population to as $|f_{M}\rangle$.  We assume that the $M$ degenerate states cannot be directly coupled with the initial state and that
 there exist $N$ nondegenerate intermediate states (denoted $|i_{1}\rangle$,  $|i_{2}\rangle$, $\cdots$, $|i_{N}\rangle$)
that can be coupled with the initial
state and some of  the
$M$ degenerate states.   The question under consideration
is the following: under what conditions can we, in principle, realize 100\% population transfer from state $|0\rangle$ 
to state $|f_{M}\rangle$ without ever populating the intermediate states or any other degenerate states. 
From a mathematical point of view, this question is formidable  because (i) the extent of  controllability in degenerate systems without constraints
is not yet resolved
and (ii) the extent of controllability in systems
subject to strong constraints has barely been studied.

\ \
\vspace{1cm}
\begin{figure}[ht]
\begin{center}
\epsfig{file=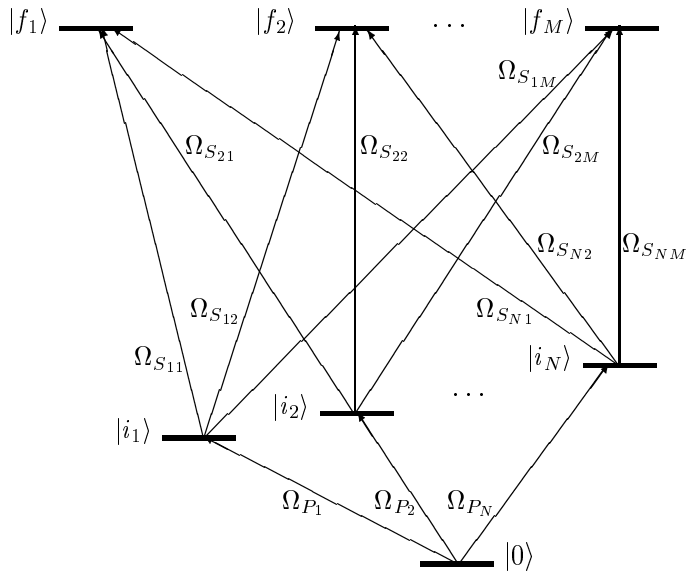,width=8cm}
\end{center}
\caption{A schematic diagram of an adiabatic passage scheme for the realization of
complete population transfer from an initial state $|0\rangle$, via $N$ nondegenerate intermediate states
 $|i_{1}\rangle$, $|i_{2}\rangle$, $\cdots$, $|i_{N}\rangle$, to state  $|f_{M}\rangle$ that is degenerate with
other $(M-1)$ states $|f_{1}\rangle$, $|f_{2}\rangle$, $\cdots$, $|f_{M-1}\rangle$.
Under certain conditions 100\% population can be transferred from
state $|0\rangle$ to state $|f_{M}\rangle$ without ever populating any of the intermediate states or any of the undesired
degenerate states.
}
\label{fig1}
\end{figure}

One needs $N$ laser fields to resonantly couple the initial state to the $N$ intermediate states, with the corresponding Rabi frequencies denoted by
$2\Omega_{P_{1}}$, $2\Omega_{P_{2}}$, $\cdots$,  $2\Omega_{P_{N}}$, and an additional $N$ laser fields
to resonantly  couple the intermediate states with the $M$ degenerate states,  with the Rabi frequency associated with state $|i_{k}\rangle$ ($k=1,2, \cdots, N$) and
state $|f_{j}\rangle$ ($j=1,2,\cdots,M$) represented by $2\Omega_{S_{kj}}$.
A schematic diagram of the energy levels and the corresponding parameters for the Rabi frequencies is shown in Fig. 1.
As in the simplest version of STIRAP \cite{bergmann},
we assume that the laser fields are Gaussian pulses and that they
are counter-intuitively ordered. Specifically,
the electric field $E_{P_{k}}$ that couples state $|0\rangle$ with state $|i_{k}\rangle$ is given by  
\beq
E_{P_{k}}(t)=\tilde{E}_{P_{k}}\cos(\omega_{P_{k}}t+\phi_{P_{k}})\exp\left[-(t-T)^{2}/T^{2}\right],  
\label{e1}
\eeq
and the electric field $E_{S_{k}}$ that couples state $|i_{k}\rangle $ with the $M$  degenerate states is given by
\beq
E_{S_{k}}(t)=\tilde{E}_{S_{k}}\cos(\omega_{S_{k}}t+\phi_{S_{k}})\exp\left[-t^{2}/T^{2}\right].
\eeq
Here $t$ is the time variable, $T$ is the pulse width, $\tilde{E}_{P_{k}}$ and $\tilde{E}_{S_{k}}$ are the peak amplitudes of the electric fields,
$\omega_{P_{k}}$ and $\omega_{S_{k}}$ are the laser carrier frequencies, 
$\phi_{P_{k}}$ and $\phi_{S_{k}}$ are the laser phases. Note that $E_{P_{k}}(t)$ is delayed by $T$ relative to $E_{S_{k}}(t)$.
The Rabi frequencies are then given by
\beq
2\Omega_{P_{k}}=2\tilde{\Omega}_{P_{k}}\exp\left[-(t-T)^{2}/T^{2}\right],
\label{peq}
\eeq
\beq
2\Omega_{S_{kj}}=2\tilde{\Omega}_{S_{kj}}\exp\left[-t^{2}/T^{2}\right],
\label{seq}
\eeq
where the peak Rabi  frequencies $2\tilde{\Omega}_{P_{k}}$ and $2\tilde{\Omega}_{S_{kj}}$ are 
\beq
2\tilde{\Omega}_{P_{k}}=\mu_{0k} \tilde{E}_{P_{k}}\exp(i\phi_{P_{k}}),
\label{peq2}
\eeq
\beq
2\tilde{\Omega}_{S_{kj}}=\mu_{kj} \tilde{E}_{S_{k}}\exp(i\phi_{S_{k}}).
\label{seq2}
\eeq
Here $\mu_{0k}\equiv \langle 0| \hat{\mu}|i_{k}\rangle$ and
 $\mu_{kj}\equiv \langle i_{k}|\hat{\mu}|f_{j}\rangle$ are the transition dipole moments. 

In the rotating-wave approximation and in the interaction representation, the Hamiltonian of the system plus the component fields of the full control field
is given by 
\begin{equation}
{\bf H}=\left[  
\begin{array}{ccccccccc}
0 & \Omega_{P_{1}} & \Omega_{P_{2}}  & \cdots & \Omega_{P_{N}} &  0 & 0 &\cdots & 0  \\
\Omega_{P_{1}}^{*} & 0 & 0 &  \cdots & 0 & \Omega_{S_{11}} & \Omega_{S_{12}} & \cdots &  \Omega_{S_{1M}}   \\
\Omega_{P_{2}}^{*} & 0 & 0 &  \cdots & 0 & \Omega_{S_{21}} & \Omega_{S_{22}} & \cdots &  \Omega_{S_{2M}}   \\
\vdots & \vdots & \vdots & \vdots &\vdots &\vdots &\vdots  & \vdots & \vdots   \\
\Omega_{P_{N}}^{*} & 0 & 0 &  \cdots & 0 & \Omega_{S_{N1}} & \Omega_{S_{N2}} & \cdots &  \Omega_{S_{NM}}   \\
0 & \Omega_{S_{11}}^{*} & \Omega_{S_{21}}^{*}  & \cdots & \Omega_{S_{N1}}^{*} &  0 & 0 &\cdots & 0  \\
0 & \Omega_{S_{12}}^{*} & \Omega_{S_{22}}^{*}  & \cdots & \Omega_{S_{N2}}^{*} &  0 & 0 &\cdots & 0  \\
\vdots & \vdots & \vdots & \vdots &\vdots &\vdots &\vdots  & \vdots & \vdots   \\
0 & \Omega_{S_{1M}}^{*} & \Omega_{S_{2M}}^{*}  & \cdots & \Omega_{S_{NM}}^{*} &  0 & 0 &\cdots & 0  \\
\end{array}
\right].
\label{hami}
\end{equation}
As suggested by this Hamiltonian, we have assumed that the $N$ nondegenerate
intermediate states will not be coupled to one another by the external fields.

\section{Adiabatic passage in degenerate systems}

Our previous work on complete control of the population transfer from
an initial state to one of two
degenerate target states \cite{gongjcp4} suggests that
the desired complete population transfer from the initial state $|0\rangle$
to the target
state  $|f_{M}\rangle$ may be achieved by
using  an  eigenstate
dressed by the external fields  that has nodes on all the states that we do not wish to transfer population to.
If, by choosing the component fields of the full control field
so that a multi-node dressed eigenstate is created that 
correlates first with state  $|0\rangle$  and  then with the  target state $|f_{M}\rangle$ as the fields evolve in time,
then in the adiabatic limit, {\it i.e.}, in cases where the dynamics of the system can adiabatically follow the evolution of 
the multi-node dressed eigenstate,
this control field will generate complete  population transfer from state  $|0\rangle$ to state $|f_{M}\rangle$.

In particular, we examine here the existence and properties of the null eigenstate (denoted $|\Lambda\rangle$) of ${\bf H}$.  By definition, one has
\beq
{\bf H}|\Lambda\rangle =0.
\label{eigeneq}
\eeq
In the same representation of ${\bf H}$,  $|\Lambda\rangle$  can be written in terms of its $(1+N+M)$ components, {\it i.e.},
$(z_{0}; x_{1}, x_{2}, \cdots, x_{N}; y_{1}, y_{2}, \cdots, y_{M})^{T}$.
Then Eqs. (\ref{hami}) and (\ref{eigeneq}) give
\beq
\sum_{k=1}^{N}\Omega_{P_{k}}x_{k}=0,
\label{eq11}
\eeq
\beq
\left(
\begin{array}{cccc}
\Omega_{S_{11}}^{*} & \Omega_{S_{21}}^{*} & \cdots &  \Omega_{S_{N1}}^{*}   \\
\Omega_{S_{12}}^{*} & \Omega_{S_{22}}^{*} & \cdots &  \Omega_{S_{N2}}^{*}   \\
\vdots & \vdots & \vdots & \vdots  \\
\Omega_{S_{1M}}^{*} & \Omega_{S_{2M}}^{*} & \cdots &  \Omega_{S_{NM}}^{*}   \\
\end{array}
\right)\left(\begin{array}{c}
x_{1}\\ x_{2} \\ \vdots \\x_{N} \\
\end{array}
\right)=
\left(\begin{array}{c}
0 \\ 0 \\ \vdots \\ 0 \\
\end{array} \right),
\label{eq22}
\eeq
and
\beq
\left(
\begin{array}{cccc}
\Omega_{S_{11}} & \Omega_{S_{12}} & \cdots &  \Omega_{S_{1M}}   \\
\Omega_{S_{21}} & \Omega_{S_{22}} & \cdots &  \Omega_{S_{2M}}   \\
\vdots & \vdots & \vdots & \vdots  \\
\Omega_{S_{N1}} & \Omega_{S_{N2}} & \cdots &  \Omega_{S_{NM}}   \\
\end{array}
\right)\left(
\begin{array}{c}
y_{1}\\ y_{2} \\  \vdots \\ y_{M} \\
\end{array}
\right)=
-z_{0}\left(\begin{array}{c} \Omega^{*}_{P_{1}}\\ \Omega^{*}_{P_{2}} \\ \vdots \\ \Omega^{*}_{P_{N}} \\ \end{array}\right).
\label{eq33}
\eeq
Below we examine the solution to Eqs. (\ref{eq11}), (\ref{eq22}) and (\ref{eq33}) in three different cases, specifically,  $M=N$, $M<N$, and $M>N$.

\subsection{The $M=N$ case}
Let $M=N$. 
We denote the $M\times M$ matrix $S_{kj}$ ($k=1,2, \cdots, M$, $j=1,2,\cdots, M$) by ${\bf S}$,
\beq
{\bf S}\equiv \left[
\begin{array}{cccc}
\Omega_{S_{11}} & \Omega_{S_{12}} & \cdots &  \Omega_{S_{1M}}   \\
\Omega_{S_{21}} & \Omega_{S_{22}} & \cdots &  \Omega_{S_{2M}}   \\
\vdots & \vdots & \vdots & \vdots  \\
\Omega_{S_{M1}} & \Omega_{S_{M2}} & \cdots &  \Omega_{S_{MM}}   \\
\end{array}
\right].
\label{sdefine}
\eeq          
Clearly,  if
\beq
\det({\bf S})\neq 0,  
\label{dnonzero}
\eeq
then  Eq. (\ref{eq22}) has only one solution 
\beq
x_{1}=x_{2}=\cdots=x_{N}=0.
\label{xeq}
\eeq
With this solution for $x_{k}$ ($k=1,2, \cdots, N$), Eq. (\ref{eq11}) is 
automatically satisfied.  Further, under condition (\ref{dnonzero}), one obtains the only
solution to Eq. (\ref{eq33}), namely 
\beq
\left(\begin{array}{c} y_{1}\\ y_{2} \\ \vdots \\ y_{M} \\ \end{array}\right)
=-z_{0}{\bf S}^{-1}\left(\begin{array}{c} \Omega^{*}_{P_{1}}\\ \Omega^{*}_{P_{2}} \\ \vdots \\ \Omega^{*}_{P_{N}} \\ \end{array}\right),
\label{yeq}
\eeq
where the value of $z_{0}$ is determined by the normalization requirement and
${\bf S}^{-1}$ denotes the inverse of ${\bf S}$. Note that ${\bf S}^{-1}$ can be explicitly written as follows:
\beq
{\bf S}^{-1}=\frac{1}{\det({\bf S})}
\left(\begin{array}{cccc}
A_{11} &  A_{21} & \cdots &  A_{M1} \\
A_{12} &  A_{22} & \cdots &  A_{M2} \\  
\vdots & \vdots & \vdots & \vdots  \\
A_{1M} & A_{2M} & \cdots &  A_{MM} \\
\end{array}
\right),
\eeq
where $A_{kj}$ is the cofactor of $\Omega_{S_{kj}}$,
\beq
A_{kj}=(-1)^{k+j}\Delta_{kj},
\eeq 
with $\Delta_{kj}$ being the complementary minor of $\Omega_{S_{kj}}$, {\it i.e.},  the determinant of an $(M-1) \times (M-1)$ submatrix of
${\bf S}$ obtained by deleting the $k$th row and $j$th column of ${\bf S}$.

It should be pointed out that whether or not  $\det({\bf S})$ can be zero is a time-independent
property inherent to the system itself.  Using Eqs. (\ref{seq}) and (\ref{seq2}) one obtains
\beq
\det({\bf S})&=&(\frac{1}{2})^{M}\left[\prod_{k=1}^{M}\tilde{E}_{S_{k}}\exp(i\phi_{S_{k}})\right]\exp(-M t^{2}/T^{2}) \nonumber \\
&&  \times
  \det 
\left(\begin{array}{cccc}
\mu_{11} & \mu_{12} & \cdots &  \mu_{1M}   \\
\mu_{21} & \mu_{22} & \cdots &  \mu_{2M}   \\
\vdots & \vdots & \vdots & \vdots  \\
\mu_{M1} & \mu_{M2} & \cdots &  \mu_{MM}   \\
\end{array}
\right).
\label{sysprop}
\eeq
Thus, provided that
\beq
\det 
\left(\begin{array}{cccc}
\mu_{11} & \mu_{12} & \cdots &  \mu_{1M}   \\
\mu_{21} & \mu_{22} & \cdots &  \mu_{2M}   \\
\vdots & \vdots & \vdots & \vdots  \\
\mu_{M1} & \mu_{M2} & \cdots &  \mu_{MM}   \\
\end{array}
\right)\neq 0,
\label{munonzero}
\eeq
$\det({\bf S})$  will be nonzero at all times. Unless there are 
some particular system symmetries, it is rare that
condition (\ref{munonzero}) will be violated.

Thus, under condition (\ref{munonzero}),
${\bf H}$ has only one null eigenvector with $N$ nodes on the 
intermediate states and its other $(M+1)$ components $z_{0}$, $y_{1}$, $y_{2}$, $\cdots$, $y_{M}$
satisfying
Eq. (\ref{yeq}). 
Due to the ordering of the laser fields [see Eqs. (\ref{peq}) and (\ref{seq})], 
this null eigenvector initially correlates with state $|0\rangle$ and then correlates with a superposition
of the $M$ degenerate states, $\sum_{k=1}^{M}y_{k}|f_{k}\rangle$.
So if the system
remains in this null eigenstate, the population 
will be transferred from state $|0\rangle$ 
to state $\sum_{k=1}^{M}y_{k}|f_{k}\rangle$ without populating any of the $N$ intermediate states as time passes.
We must still examine whether or not
it is possible to realize complete population transfer to one arbitrary memeber of the $M$ degenerate states without ever populating
the other $(M-1)$ degenerate states, and if yes,
under what conditions.

Our consideration is based on the relation
\beq
A_{1k}\Omega_{S_{1M}}+A_{2k}\Omega_{S_{2M}} + \cdots + A_{Nk}\Omega_{S_{NM}} =\delta_{kM}\det({\bf S}),
\eeq
which leads to
\beq
{\bf S}^{-1}\left(\begin{array}{c}
\Omega_{S_{1M}}\\
\Omega_{S_{2M}}\\
\vdots\\
\Omega_{S_{NM}}\\
\end{array} \right) = \left(\begin{array}{c}0 \\ 0  \\ \vdots \\ 1 \end{array}\right).
\label{zeroeq}
\eeq
Hence, if the laser fields are designed such that
\beq
\left(\begin{array}{c}
\Omega_{P_{1}}^{*}\\
\Omega_{P_{2}}^{*}\\
\vdots\\
\Omega_{P_{N}}^{*}\\ \end{array} \right)=
\xi(t)
\left(\begin{array}{c}
\Omega_{S_{1M}}\\
\Omega_{S_{2M}}\\
\vdots\\
\Omega_{S_{NM}}\\ \end{array}
\right),
\label{keycon}
\eeq
where $\xi(t)$ is any function of time,  
then Eqs. (\ref{yeq}), (\ref{zeroeq}) and (\ref{keycon}) yield
\beq
\left(\begin{array}{c} y_{1}\\ y_{2} \\ \vdots \\ y_{M} \\ \end{array}\right)
=-z_{0}\xi(t)\left(\begin{array}{c}0 \\ 0  \\ \vdots \\ 1 \end{array}\right).
\label{s2}
\eeq
That is, by manipulating the component fields of the total control field
we can guarantee that
the only null eigenvector of ${\bf H}$ 
will have additional $(M-1)$ nodes on states
$|f_{1}\rangle$, $|f_{2}\rangle$, $\cdots$, $|f_{M-1}\rangle$.  
The total number of nodes of $|\Lambda\rangle$ is then given by $(M+N-1)$.
It follows that if the dynamics of population transfer adiabatically follows the 
evolution of $|\Lambda\rangle$, 
complete population transfer from state $|0\rangle$ to one arbitrary member of the $M$
degenerate states can be achieved without ever populating the other
$(M-1)$ degenerate states or the $N$ intermediate states.

For the specific pulse shape functions given in Eqs. (\ref{peq}) and (\ref{seq}), condition (\ref{keycon}) is equivalent to
\beq
\left(
\begin{array}{c}
\tilde{\Omega}^{*}_{P_{1}}\\
\tilde{\Omega}^{*}_{P_{2}}\\
\vdots
\\
\tilde{\Omega}^{*}_{P_{N}}\\
\end{array}
\right)
=\eta
\left(
\begin{array}{c}
\tilde{\Omega}_{S_{1M}} \\
\tilde{\Omega}_{S_{2M}} \\
\vdots \\
\tilde{\Omega}_{S_{NM}}\\ \end{array} \right),
\label{keycon0}
\eeq
where $\eta$ is a nonzero constant.
Using Eqs. (\ref{peq2}) and (\ref{seq2}), the above condition can be
translated to
\beq
\left(
\begin{array}{c}
\mu_{01}^{*}\tilde{E}_{P_{1}}\exp(-i\phi_{P_{1}})\\
\mu_{02}^{*}\tilde{E}_{P_{2}}\exp(-i\phi_{P_{2}})\\
\vdots \\
\mu_{0N}^{*}\tilde{E}_{P_{N}}\exp(-i\phi_{P_{N}}) \\ \end{array}\right)
=\eta \left(
\begin{array}{c}
\mu_{1M}\tilde{E}_{S_{1}}\exp(i\phi_{S_{1}})\\
\mu_{2M}\tilde{E}_{S_{2}}\exp(i\phi_{S_{2}})\\
\vdots \\
\mu_{NM}\tilde{E}_{S_{N}}\exp(i\phi_{S_{N}})\\ \end{array} \right).
\label{keycon1}
\eeq
Surprisingly, Eq. (\ref{keycon1}) makes no reference to the transition dipole moments that are related to
states $|f_{1}\rangle$, $|f_{2}\rangle$, $\cdots$, $|f_{M-1}\rangle$. Then,
 even if
one has no knowledge about the properties of states $|f_{1}\rangle$, $|f_{2}\rangle$, $\cdots$, $|f_{M-1}\rangle$ beforehand,
it is still possible, by use of adiabatic passage,
to completely suppress the population transferred to these states. Note also that 
condition (\ref{keycon1}) requires a definite phase relationship between the 
component fields of the full control field, which is a consequence
of the fact that 
quantum interference effects are directly responsible for the appearance of the additional $(M-1)$ nodes of $|\Lambda\rangle$.
 
There is another interesting implication of  Eq.  (\ref{keycon1}). Let us assume that among the $N$ transition dipole moments related to state
$|f_{M}\rangle$, $l$ of them, say, $\mu_{1M}$, $\mu_{2M}$, $\cdots$, $\mu_{lM}$,  happen to be zero. To still
have an $(M+N-1)$-node null eigenvector
of ${\bf H}$,
Eq.  (\ref{keycon1}) requires $\tilde{E}_{P_{1}}=\tilde{E}_{P_{2}}=\cdots=\tilde{E}_{P_{l}}=0$, {\it i.e.}, removing the $l$ laser fields
that couple state $|0\rangle$ to the intermediate states $|i_{1}\rangle$, $|i_{2}\rangle$, $\cdots$,  $|i_{l}\rangle$.
This clearly results in a simplification of our adiabatic passage scheme, but
it does not suggest that in these cases the intermediate states $|i_{1}\rangle$,  $|i_{2}\rangle$, $\cdots$,  $|i_{l}\rangle$  are no longer important or
can be removed from the system,
because they are still coupled
with the other unwanted degenerate states. These intermediate states in this situation can be regarded as 
a high-dimensional analog of the so-called branch state
in a five-level extended STIRAP scheme 
\cite{kobrakpra}, for which case it has been shown
that increasing  the coupling between an unwanted state and the branch state
will enhance the yield of the desired state. 

\subsection{The $ M<N$ case}
In this case Eq. (\ref{eq33}) requires
\beq
{\bf S}\left(\begin{array}{c}y_{1} \\ y_{2} \\ \vdots \\ y_{M} \\
\end{array}\right)= -z_{0}\left(\begin{array}{c} \Omega^{*}_{P_{1}}\\ \Omega^{*}_{P_{2}} \\ \vdots \\ \Omega^{*}_{P_{N}} \\ \end{array}\right),
\label{eq332}
\eeq
and
\beq
\sum_{j=1}^{M}\Omega_{S_{kj}}y_{j}=-z_{0}\Omega_{P_{k}}^{*}, \ k=M+1, M+2, \cdots, N.
\label{mn2}
\eeq
Under condition (\ref{keycon}), Eq. (\ref{eq332}) still gives the result of Eq. (\ref{s2}), which can also ensure that 
Eq. (\ref{mn2}) is satisfied.  It is then  clear that the null eigenvector, if 
it exists, will still have $(M-1)$ nodes 
on the unwanted degenerate states.

Next we examine Eqs. (\ref{eq11}) and (\ref{eq22}) for $M<N$.  Note first that 
under condition (\ref{keycon}), Eq. (\ref{eq11}) 
is included in Eq. (\ref{eq22}).   
Equation (\ref{eq22}) indicates that now there are $N$ variables $x_{1}$, $x_{2}$, $\cdots$, $x_{N}$
and yet the number of the constraints is only $M$.  Hence there are more than one solutions to  Eq. (\ref{eq22}), {\it i.e.},
${\bf H}$ can have multiple null eigenvectors if $M<N$.

The most obvious solution to Eq. (\ref{eq22}) is given by $x_{1}=x_{2}=\cdots=x_{N}=0$.  Under condition (\ref{keycon})
the associated null eigenvector
is then given by 
\beq
|\Lambda\rangle=|\Lambda_{1}\rangle\equiv (z_{0};0,0,\cdots, 0; 0,0,\cdots, -z_{0}\xi(t))^{T}.
\label{null1}
\eeq
This eigenvector is analogous to the only null eigenvector 
in the $M=N$ case. Let the
other null eigenvectors be denoted 
$|\Lambda_{2}\rangle$. Under condition (\ref{keycon}) 
$|\Lambda_{2}\rangle$
can be written as
\beq
|\Lambda_{2}\rangle\equiv (z_{0};x_{1},x_{2},\cdots, x_{N}; 0,0,\cdots, -z_{0}\xi(t))^{T}, 
\label{othernull}
\eeq
with $x_{1},x_{2},\cdots, x_{N}$ satisfying Eq. (\ref{eq22}) and at least one of them being nonzero.  
Since any linear combination of $|\Lambda_{1}\rangle$ and $|\Lambda_{2}\rangle$ is still a null eigenvector,
we rewrite the other null eigenvectors as 
\beq
|\Lambda_{3}\rangle & \equiv & |\Lambda_{2}\rangle-|\Lambda_{1}\rangle \nonumber \\
&=&  (0;x_{1},x_{2},\cdots, x_{N}; 0,0,\cdots, 0)^{T}.
\label{othernull2}
\eeq
Note that $|\Lambda_{3}\rangle$ is orthogonal to $\Lambda_{1}$. As such,
at early times state $|\Lambda_{1}\rangle$ fully correlates with the initial state $|0\rangle$, but all the other
null eigenvectors represented by $|\Lambda_{3}\rangle$ have zero overlap with state $|0\rangle$.  

The important question is whether or not one can avoid populating states $|\Lambda_{3}\rangle$ and therefore 
avoid populating the intermediate states during the population transfer.
At first glance it seems plausible that since states $|\Lambda_{3}\rangle$ are degenerate with  $|\Lambda_{1}\rangle$, nonadiabatic transitions
from $|\Lambda_{1}\rangle$ to  $|\Lambda_{3}\rangle$ will be unavoidable. Interestingly,  this is not the case.
Specifically, 
the strength of the nonadiabatic coupling between $|\Lambda_{3}\rangle$ and $|\Lambda_{1}\rangle$
is proportional to $|\langle \Lambda_{3}| d  \Lambda_{1}/ dt \rangle|$, which turns out to be
zero at all  times 
due to the $(M+N-1)$-node
structure of $|\Lambda_{1}\rangle$ and the $(M+1)$-node structure of $|\Lambda_{3}\rangle$
[see Eqs. (\ref{null1}) and (\ref{othernull2})].
Hence, only the null eigenvector $|\Lambda_{1}\rangle$ that has $(M+N-1)$
nodes is relevant to the dynamics of population transfer.  

The analysis here leads to a significant prediction. That is, under condition (\ref{keycon1}) and for $M\leq N$, 
we can always achieve adiabatic passage via the null eigenvector
$|\Lambda_{1}\rangle$, irrespective of the actual value of $M$ and the values of
the transition dipole moments between the intermediate states
and the unwanted degenerate states.
As a result, during the entire process of
population transfer the undesired degenerate 
states and all the intermediate states
can be never populated,
even if the exact number of the unwanted degenerate states and 
their properties are unknown to us. 

Of course, if the exact number of degeneracies $M<N$ is given,  it is unnecessary
to still use the $2N$ component fields to achieve complete population transfer. As indicated in the preceding subsection,
one may remove $2(N-M)$ laser fields and consider 
only $M$ intermediate states
to realize the same control.

\subsection{The $M>N$ case}

In this case Eqs. (\ref{eq11}) and (\ref{eq22}) in general have only one solution given by Eq. (\ref{xeq}).
So the null eigenvector, if it exists, should generically have $N$ nodes on all the intermediate states.
However, there are multiple solutions to Eq. (\ref{eq33}) since there are $M$ unknown variables $y_{1}$, $y_{2}$, $\cdots$,
$y_{M}$ whilst the number of the constraints is only $N$. Therefore ${\bf H}$ again has multiple null eigenvectors.
Denote two orthogonal 
null eigenvectors of ${\bf H}$ by $|\Lambda'\rangle$ and $|\Lambda''\rangle$,
\beq
|\Lambda'\rangle& = &(z_{0}';0,0,\cdots,0; y_{1}', y_{2}', \cdots, y_{M}')^{T}, \nonumber \\
|\Lambda''\rangle& = &(z_{0}'';0,0,\cdots,0; y_{1}'', y_{2}'', \cdots, y_{M}'')^{T}.
\eeq
The strength (denoted $\chi$) of the nonadiabatic coupling between states $|\Lambda'\rangle$ and $|\Lambda''\rangle$ is proportional to
$|(z_{0}')^{*}dz_{0}''/dt + \sum_{j=1}^{M}(y_{j}')^{*}dy_{j}''/dt|$. 
Due to the normalization requirement,
$z_{0}'=y_{1}'=y_{2}'= \cdots=y_{M}'=0$ is not an acceptable solution to  Eq. (\ref{eq33}). As such, there is no general reason why
$\chi$ should be zero.  In particular, under condition (\ref{keycon}), one finds that one null eigenvector
can be analogous to that in the $M\leq N$ case, {\it i.e.},
$|\Lambda'\rangle = (z_{0};0,0,\cdots,0; 0, 0, \cdots, -z_{0}\xi(t))^{T}$. Then 
$\chi \propto |z_{0}^{*}dz_{0}''/dt - z_{0}^{*}\xi^{*}(t)dy_{M}''/dt|$, which is nonzero in general. 
This being the case,
state $|\Lambda''\rangle$ and therefore some of the states $|f_{1}\rangle$, $|f_{2}\rangle$, $\cdots$, $|f_{M-1}\rangle$
will be populated during the population transfer. This is true no matter how slowly the laser fields
are turned on or off, as states $|\Lambda''\rangle$ and  $|\Lambda'\rangle$ are degenerate.

Clearly, then, in the $M>N$ case, our scheme cannot guarantee complete  population transfer to state $|f_{M}\rangle$.  However,
the results above suggest that 
this problem can be easily fixed if there are more intermediate states available. That is, using  additional $(M-N)$ intermediate states
and adding $2(M-N)$ laser fields will recover the desired complete population transfer.

\subsection{Complete population transfer to arbitrary superpositions of $M$ degenerate states}
We note that states $|f_{1}\rangle$, $|f_{2}\rangle$, $\cdots$, $|f_{M}\rangle$ are
just one particular choice of the basis states of an $M$-dimensional degenerate subspace.  Indeed, any arbitrary superposition
of states  $|f_{1}\rangle$, $|f_{2}\rangle$, $\cdots$, $|f_{M}\rangle$ can be used as one of the basis states of the same
degenerate subspace.  Consider now a target superposition state 
\beq
|f_{M}'\rangle=\sum_{j=1}^{M}c_{j}|f_{j}\rangle,
\eeq
where $c_{j}$ ($j=1,2, \cdots, M)$ are arbitrary coefficients in the superposition state. 
Let $|f_{M}'\rangle$ be the last new basis state of the same degenerate subspace and let  the other 
new orthogonal basis states be $|f_{1}'\rangle$, $|f_{2}'\rangle$, $\cdots$, $|f_{M-1}'\rangle$.  
The transition dipole moment between state $|i_{k}\rangle$ and
 state $|f_{j}'\rangle$ is represented by $\mu_{kj}'$.

We now apply the above general solution to complete population transfer in degenerate systems to
state $|f_{M}'\rangle$ instead of
$|f_{M}\rangle$.  Then,  for $M\leq N$ and
\beq
\det
\left(\begin{array}{cccc}
\mu_{11}' & \mu_{12}' & \cdots &  \mu_{1M}'   \\
\mu_{21}' & \mu_{22}' & \cdots &  \mu_{2M}'   \\
\vdots & \vdots & \vdots & \vdots  \\
\mu_{M1}' & \mu_{M2}' & \cdots &  \mu_{MM}'   \\
\end{array}
\right)\neq 0,
\label{munonzero2}
\eeq
all population can be transferred from state $|0\rangle$ to the superposition state $|f_{M}'\rangle$
if the component fields
satisfy
\beq
\left(
\begin{array}{c}
\mu_{01}^{*}\tilde{E}_{P_{1}}\exp(-i\phi_{P_{1}})\\
\mu_{02}^{*}\tilde{E}_{P_{2}}\exp(-i\phi_{P_{2}})\\
\vdots \\
\mu_{0N}^{*}\tilde{E}_{P_{N}}\exp(-i\phi_{P_{N}}) \\ \end{array}\right)
=\eta \left(
\begin{array}{c}
\mu_{1M}'\tilde{E}_{S_{1}}\exp(i\phi_{S_{1}})\\
\mu_{2M}'\tilde{E}_{S_{2}}\exp(i\phi_{S_{2}})\\
\vdots \\
\mu_{NM}'\tilde{E}_{S_{N}}\exp(i\phi_{S_{N}})\\ \end{array} \right),
\label{keycon3}
\eeq
where
\beq
\mu_{kM}'=\sum_{j=1}^{M}c_{j}\mu_{kj}, \ k=1,2, \cdots, N.
\eeq
Since the transition dipole moments $\mu_{kj}'$ can be obtained from $\mu_{kj}$ by considering a unitary transformation
between two different sets of basis states of the same degenerate
subspace, it can be easily proved that condition (\ref{munonzero2}) is equivalent to condition (\ref{munonzero}).
Note also that there is no need to construct,  from superpositions of
states $|f_{1}\rangle$, $|f_{2}\rangle$, $\cdots$, $|f_{M}\rangle$,
the explicit forms of states $|f_{1}'\rangle$, $|f_{2}'\rangle$, $\cdots$, $|f_{M-1}'\rangle$,
as their properties are not required to predict that 
they can be never populated during the population transfer.

\section{Numerical Example}

To confirm the validity and feasibility of our general solution to the realization of complete population transfer to one arbitrary member of a set of
$M$ degenerate states,
we present in this section some numerical examples.  
 In particular, we consider a system that has $N=7$ nondegenerate states as intermediate states
and $M\leq N$ degenerate orthogonal states.  We define the sum of population in all the
intermediate states as $P_{x}$, the sum of population in
all the degenerate states except for state $|f_{M}\rangle$ as $P_{y}$, and the population in the target state $|f_{7}\rangle$ as $P_{f}$.
For convenience all the transition dipole moments are assumed to be real and all the laser phases are set to zero.  Hence all the
Rabi frequencies
take real values. To fulfill condition (\ref{keycon0}), we assume $\tilde{\Omega}_{P_{k}}=\tilde{\Omega}_{S_{k7}}$, for $k=1,2,\cdots, N$.

\begin{figure}[ht]
\begin{center}
\epsfig{file=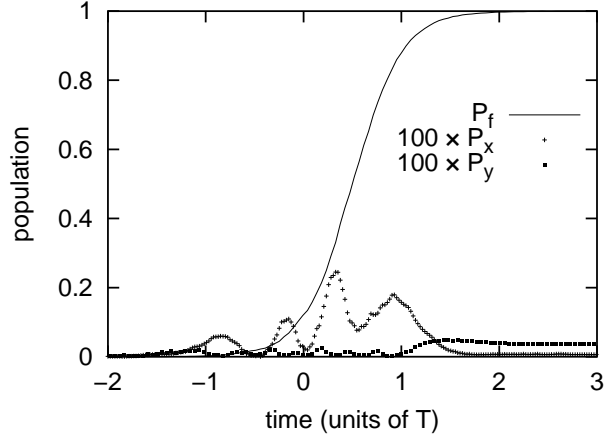,width=6cm,angle=270}
\end{center}
\caption{A numerical example of adiabatic passage through a 13-node null eigenstate
for the complete population transfer from an initial state
to one arbitrary member of $M=7$
degenerate states. The number of intermediate states is given by $N=7$.
 $P_{x}$ represents the population in all the intermediate states,  $P_{y}$ represents the population
in all the degenerate states except for the target state, and $P_{f}$ represents the population in the target state $|f_{7}\rangle$.
The parameters for
the peak Rabi frequencies are, in units of $1/T$,
given by ($\tilde{\Omega}_{P_{1}}$, $\tilde{\Omega}_{P_{2}}$, $\tilde{\Omega}_{P_{3}}$,
$\tilde{\Omega}_{P_{4}},\tilde{\Omega}_{P_{5}}$, $\tilde{\Omega}_{P_{6}}$, $\tilde{\Omega}_{P_{7}}) =$ (60, 90, 60, 120, 90, 99, 135),
($\tilde{\Omega}_{S_{11}}$, $\tilde{\Omega}_{S_{12}}$, $\tilde{\Omega}_{S_{13}}$, $\tilde{\Omega}_{S_{14}}$, $\tilde{\Omega}_{S_{15}}$,
$\tilde{\Omega}_{S_{16}}$, $\tilde{\Omega}_{S_{17}})  = $
(90, 15, 0, 150, 36, 18, 60),
($\tilde{\Omega}_{S_{21}}$, $\tilde{\Omega}_{S_{22}}$, $\tilde{\Omega}_{S_{23}}$, $\tilde{\Omega}_{S_{24}}$, $
\tilde{\Omega}_{S_{25}}$, $\tilde{\Omega}_{S_{26}}$, $\tilde{\Omega}_{S_{27}})= $
(90, 57, 24, 45, 69, 78, 90),
($\tilde{\Omega}_{S_{31}}$, $\tilde{\Omega}_{S_{32}}$, $\tilde{\Omega}_{S_{33}}$, $\tilde{\Omega}_{S_{34}}$, $
\tilde{\Omega}_{S_{35}}$, $\tilde{\Omega}_{S_{36}}$, $\tilde{\Omega}_{S_{37}})= $
(90, 75, 39, 36, 39, 78, 60),
($\tilde{\Omega}_{S_{41}}$, $\tilde{\Omega}_{S_{42}}$, $\tilde{\Omega}_{S_{43}}$, $\tilde{\Omega}_{S_{44}}$, $
\tilde{\Omega}_{S_{45}}$, $\tilde{\Omega}_{S_{46}}$, $\tilde{\Omega}_{S_{47}})= $
(60, 18, 24, 75, 66, 48, 120),
($\tilde{\Omega}_{S_{51}}$, $\tilde{\Omega}_{S_{52}}$, $\tilde{\Omega}_{S_{53}}$, $\tilde{\Omega}_{S_{54}}$, $
\tilde{\Omega}_{S_{55}}$, $\tilde{\Omega}_{S_{56}}$, $\tilde{\Omega}_{S_{57}})= $
(39, 27, 93, 15, 66, 78, 90),
($\tilde{\Omega}_{S_{61}}$, $\tilde{\Omega}_{S_{62}}$, $\tilde{\Omega}_{S_{63}}$, $\tilde{\Omega}_{S_{64}}$, $
\tilde{\Omega}_{S_{65}}$, $\tilde{\Omega}_{S_{66}}$, $\tilde{\Omega}_{S_{67}})= $
(93, 69, 18, 87, 72, 78, 99), and
($\tilde{\Omega}_{S_{71}}$, $\tilde{\Omega}_{S_{72}}$, $\tilde{\Omega}_{S_{73}}$, $\tilde{\Omega}_{S_{74}}$, $
\tilde{\Omega}_{S_{75}}$, $\tilde{\Omega}_{S_{76}}$, $\tilde{\Omega}_{S_{77}})=$
(36, 54, 48, 57, 96, 78, 135).
Note that condition (\ref{keycon0}) is satisfied since we have chosen $\tilde{\Omega}_{P_{k}}=\tilde{\Omega}_{S_{k7}}$, for $k=1,2,\cdots, 7$.
At all times $P_{x}< 0.3\%$ and $P_{y}<0.05\%$.
}
\label{fig2}
\end{figure}

Figure 2 displays an example of population transfer in the case of $M=N$. It is seen that during the population transfer, the maximum value of $P_{x}$
is less than 0.3\%, and the maximum value of $P_{y}$ is less than 0.1\%.  
The final $P_{f}$ is extremely close to 100\%.
We then arbitrarily alter some of the peak Rabi frequencies
that are related to states $|f_{1}\rangle$, $|f_{2}\rangle$, $\cdots$ $|f_{6}\rangle$. The
associated results are shown in Fig. 3.  Clearly, population transfer is still
almost complete and $P_{x}$ and $P_{y}$ again remain negligible at all times.
This demonstrates that totally suppressing the population transferred to states $|f_{1}\rangle$, $|f_{2}\rangle$, $\cdots$,
$|f_{6}\rangle$ does not require us to know their properties beforehand.  
Similar results are obtained in numerous other numerical experiments.

\begin{figure}[ht]
\begin{center}
\epsfig{file=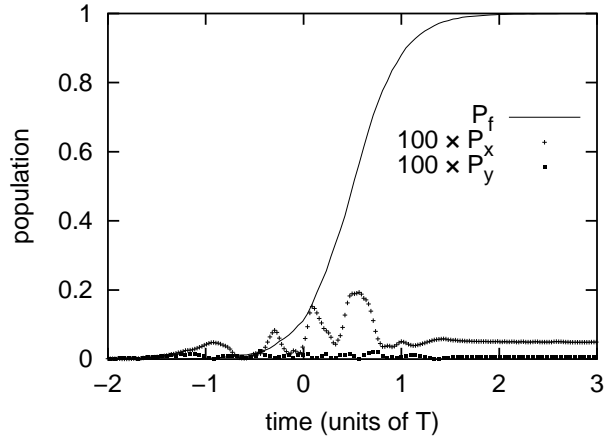,width=6cm,angle=270}
\end{center}
\caption{Same as in Fig. 2 except $\tilde{\Omega}_{S_{14}}=39$, $\tilde{\Omega}_{S_{25}}=
39$, $\tilde{\Omega}_{S_{26}}=48$,
$\tilde{\Omega}_{S_{32}}=45$, $\tilde{\Omega}_{S_{43}}=81$, $\tilde{\Omega}_{S_{
52}}=57$, $\tilde{\Omega}_{S_{56}}=48$,
 $\tilde{\Omega}_{S_{62}}=39$, $\tilde{\Omega}_{S_{72}}=24$.
At all times $P_{x}< 0.2\%$ and $P_{y}<0.03\%$.
}
\label{fig3}
\end{figure}

Figure 4 displays analogous results in a case where two transition dipole moments $\mu_{17}$ and  $\mu_{27}$
(and therefore the two peak Rabi frequencies $\tilde{\Omega}_{S_{17}}$ and $\tilde{\Omega}_{S_{27}}$)
that are related to the target state $|f_{7}\rangle$ happen to be zero.  As predicted by Eq. (\ref{keycon1}), this requires us to
remove two laser fields that connect the initial state to two intermediate states.  Accordingly we set $\tilde{\Omega}_{P_{1}}$ and
$\tilde{\Omega}_{P_{2}}$ to zero.  As seen from Fig. 4, almost
complete population transfer with $P_{x}$ and $P_{y}$ negligible at all times
is also achieved.

\begin{figure}[ht]
\begin{center}
\epsfig{file=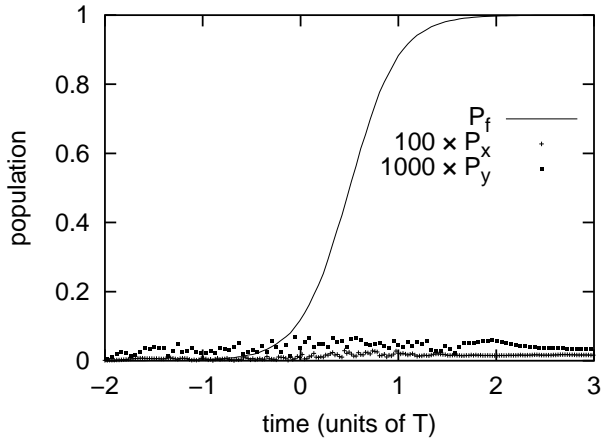,width=6cm,angle=270}
\end{center}
\caption{
Same as in Fig. 2 except $\tilde{\Omega}_{S_{17}}=\tilde{\Omega}_{S_{27}}=0$ and
$\tilde{\Omega}_{P_{1}}=\tilde{\Omega}_{P_{2}}=0$.
At all times $P_{x}< 0.03\%$ and $P_{y}<0.01\%$.}
\label{fig4}
\end{figure}

Figure 5 shows a numerical example where all the peak Rabi frequencies that are related to states  $|f_{1}\rangle$ and $|f_{2}\rangle$
are set to zero, while keeping other peak Rabi frequencies the same as those used in Fig. 2.
Due to this procedure, $|f_{1}\rangle$ and $|f_{2}\rangle$ are totally decoupled from the system and
effectively there are only five degenerate states.  As seen from Fig. 5,  almost
100\% population transfer to state
$|f_{7}\rangle$ with $P_{x}$ and $P_{y}$ negligible at all times is also obtained.
This clearly confirms our previous prediction
that our adiabatic passage method works as well even if the exact number of
degeneracies ({\it e.g.,} $M=5$ or $M=7$) is unknown to us.

\begin{figure}[ht]
\begin{center}
\epsfig{file=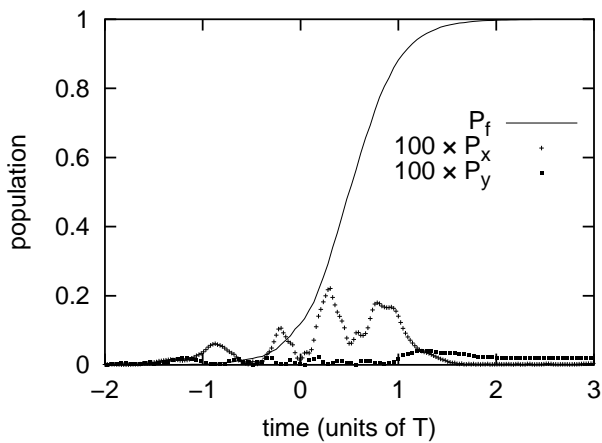,width=6cm,angle=270}
\end{center}
\caption{Same as in Fig. 2 except that all the peak Rabi frequencies that are related to
states $|f_{1}\rangle$ and $|f_{2}\rangle$ are set to zero,
{\it i.e.}, $\tilde{\Omega}_{S_{k1}}=\tilde{\Omega}_{S_{k2}}=0$, for $k=1,2,
\cdots, 7$.
At all times $P_{x}< 0.3\%$ and $P_{y}<0.05\%$.
}
\label{fig5}
\end{figure}

To end this section, we note that by increasing the pulse width or the field intensity, or optimizing the time delay between the laser fields and the ratio
of the peak Rabi frequencies (e.g., $\tilde{\Omega}_{P_{k}}/\tilde{\Omega}_{S_{kM}}$),
one can always further suppress the nonadiabaticity of the dynamics and therefore the population transfer
can be made even closer to the adiabatic limit.

\section{Discussion and Conclusion}

Controlled population transfer from a selected initial state
to a manifold of degenerate states has previously been examined 
in the context of atomic processes involving magnetic sublevels \cite{unanyan0,bergmannprl,kis,unanyan,karpati}.
Our results are much more powerful and general than those
 previously established.
A key difference between this work and previous studies is that we have exploited the properties of particular coherent excitations of manifolds with
multiple nondegenerate intermediate
states. In particular, with $N$ nondegenerate intermediate states
it becomes possible to adjust, with $2N$ laser fields, 
all the Rabi frequencies involved in Eq. (\ref{keycon}).
Hence Eq. (\ref{keycon}) can always be satisfied by designing the laser fields,
providing thereby  a systematic solution to the design of
complete population transfer from an initial state to an arbitrary superposition of $M\leq N$ degenerate states.  It is also straightforward to determine
if the requirements ({\it i.e.}, condition (\ref{munonzero}) and $M\leq N$)
are met for our general solution to be applicable.

Most coherent control approaches based on STIRAP-like dynamics are independent
of the relative phases of the control laser fields \cite{gongjcp1}.  As shown in this work
and in Refs. \cite{gongjcp4} and \cite{karpati},
this is not true for all STIRAP-like control schemes.  On one hand,
the phase-dependence of the control process
causes even more experimental difficulties; on the other hand
the phase-dependence of our adiabatic passage scheme clearly demonstrates that
laser phases can be also important in STIRAP-like dynamics and may bring about dramatic opportunities for
controlling
atomic and molecular processes.

One great advantage of our adiabatic passage scheme is its adaptability
in cases where the intermediate states or the unwanted degenerate states
are not bound states. That is, since the $(M+N-1)$-node null eigenvector
does not overlap with any of the intermediate states or any of
the unwanted degenerate states,
the structure of this multi-node eigenvector survives
if these states actually
decay.
As long as the associated decay rate constants are much
smaller than the peak Rabi frequencies of the component fields \cite{gongjcp4},
the nonadiabaticity induced by the decaying states will be negligible
and the dynamics of population transfer would be still given by the evolution
of the $(M+N-1)$-node null eigenvector.

This work extends our previous adiabatic passage method
for the realization of complete control of the population transfer branching ratio between two degenerate states \cite{gongjcp4}.
The extension is from a five-level and four-pulse scheme to a $(1+N+M)$-level and $(2N)$-pulse scheme.
While the five-level and four-pulse scheme should be experimentally achievable, 
directly applying the present work to realistic systems would be demanding for $M\geq 3$, as it would in general require 
a significant number of laser fields.
However, from the point of view of understanding the limits to
complete controllability in 
degenerate systems subject to strong constrains and/or with unknown parameters, this work is of
great theoretical interest.  In particular, we have shown that
it is possible to realize complete population transfer
without ever populating a Hilbert subspace that is only dimension-two smaller than the whole Hilbert space, even when
the exact number of degeneracies and the properties of the unwanted degenerate states are unknown to us. We hope that
the results of this study will
 motivate future theretical work on coherent control in degenerate systems.  For example,
it is interesting to consider variations on our adiabatic passage scheme, and  to
seek alternative, and  experimentally more feasible, control scenarios that can provide the same type of
population transfer pathway that is already shown to exist.

Although our adiabatic passage solution offers a general method for complete population transfer 
from an initial state to an arbitrary member of a set of
$M$
degenerate states, there are still two subtle differences between this solution and
an actual proof of complete controllability of the degenerate system. Firstly,
a completely controllable degenerate system will have
 a complete population transfer pathway
within a finite time, whereas the complete population transfer
in our solution, in the most strict sense ({\it i.e.}, $P_{f}$ is
exactly 100\%), can only be realized 
in the adiabatic limit, {\it e.g.}, for infinitely large pulse width.
Secondly, we have used the rotating-wave approximation 
 in constructing our solution.  
Mathematically speaking, such an approximation
inevitably modifies the issue of complete controllability. It is unclear to what physical extent this approximation
changes the controllability of degenerate systems.
For the above two reasons we regard our adiabatic passage method as a specific
physical, but not mathematical, solution to
the complete control of degenerate systems.

One potential application of this study is to provide a useful guide for understanding the extremely complex laser fields obtained
by genetic algorithms in adaptive feedback control of quantum systems \cite{ricebook,rice01,rabitzs}. Suppose an adapative feedback control
experiment is carried out in a degenerate system and the control goal is to maximize the population in only one of $M$ degenerate
states and minimize the population in all the unwanted degenerate
states and the
intermediate states.
It would be of great interest to
see if the optimized control field suggested by a genetic algorithm can be decomposed into $2N\geq 2M $ laser fields with their 
frequencies, amplitudes, and relative phases close to what is suggested by our adiabatic passage scheme.
If this is the case, then such a feedback control experiment becomes
a realization of our general solution to complete population transfer
in degenerate systems and more applications of this work 
can be expected.

To summarize, we have shown that under certain conditions complete population transfer
from an initial state, via $N$ nondegenerate intermediate states,
to an arbitrary superposition of $M$ ($M\leq N$) degenerate states
is achievable by adiabatically following an $(M+N-1)$-node null eigenstate that is created by designing the relative phases and
amplitudes of the component fields of a control field.
The results may find applications in quantum information processing
in atoms and molecules and 
shed considerable light on the issue of complete controllability in degenerate systems
under strong constraints and/or with unknown parameters.

\section{Acknowledgments}
This work was supported by the National Science Foundation.

\end{document}